\documentclass[10pt,twocolumn, aps,pr,showpacs,superscriptaddress,groupedaddress,nofootinbib]{revtex4}  

\usepackage{longtable}

\usepackage{ulem}   
\usepackage{comment} 
\usepackage{datetime}

\usepackage{tabularx}
\usepackage{float}
\restylefloat{table}

\usepackage[cmtip,arrow]{xy}
\usepackage{pb-diagram,pb-xy}

\usepackage{amsmath}
\usepackage{amssymb}
\usepackage{amsthm}
\usepackage[pdftex]{color}
\usepackage[pdftex,colorlinks,citecolor=blue,linkcolor=blue,urlcolor=blue]{hyperref} 
\usepackage{graphicx}
\usepackage{dcolumn} 
\usepackage{bm} 

\usepackage{longtable}

\usepackage{ulem}   
\usepackage{comment} 
\usepackage{datetime}

\usepackage{tabularx}
\usepackage{float}
\restylefloat{table}

\newcommand{\beq}{\begin{equation}}
\newcommand{\eeq}{\end{equation}}
\newcommand{\bea}{\begin{eqnarray}}
\newcommand{\eea}{\end{eqnarray}}
\newcommand{\barr}{\begin{array}}
\newcommand{\earr}{\end{array}}

\long\def\begincomment#1\endcomment{}

\newcommand{\ds}{\mathbb{d}}

\usepackage{pgf,tikz}
\usetikzlibrary{arrows}

\usepackage{bm}
\usepackage{bbold}

\usepackage{mathtools}
\DeclarePairedDelimiter\bra{\langle}{\rvert}
\DeclarePairedDelimiter\ket{\lvert}{\rangle}
\DeclarePairedDelimiterX\braket[2]{\langle}{\rangle}{#1 \delimsize\vert #2}

\begin{document}


\title{   Lattice oscillator model on noncommutative space: eigenvalues problem for the perturbation theory
}

\author{Dine Ousmane Samary}
\email{dine.ousmane.samary@aei.mpg.de}
\affiliation{International Chair in Mathematical Physics and Applications (ICMPA-UNESCO Chair), University of Abomey-Calavi,
072B.P.50, Cotonou, Republic of Benin}
\affiliation{Max Planck Institute for Gravitational Physics, Albert Einstein Institute, Am M\"uhlenberg 1, 14476, Potsdam, Germany}

\author{S\^ecloka Lazare Guedezounme} \email{guesel10@yahoo.fr}  
\affiliation{International Chair in Mathematical Physics and Applications (ICMPA-UNESCO Chair), University of Abomey-Calavi,
072B.P.50, Cotonou, Republic of Benin}

\author{Antonin Danvid\'e Kanfon} 
\email{kanfon@yahoo.fr}
\affiliation{International Chair in Mathematical Physics and Applications (ICMPA-UNESCO Chair), University of Abomey-Calavi,
072B.P.50, Cotonou, Republic of Benin}

\date{\today}

\begin{abstract}Harmonic  oscillator in noncommutative two dimensional lattice is investigated. Using the properties of non-differential calculus and its applications to quantum mechanics, we provide the eigenvalues and eigenfunctions of the corresponding Hamiltonian. First we consider the case of ordinary quantum mechanics, and we point out the thermodynamic properties of the model. Then we consider the same question when both coordinates and momenta are noncommutative.
\end{abstract}


\maketitle

\section{Introduction} \label{sec1}
Several experiments and theoretical results show that the continuous space time, in the description of modern physics, should need revision at the scale where quantum theory and gravitation can be conciliated \cite{Jizba:2011zz}-\cite{Jizba:2009qf}. The discrete spacetime has become a tool of choice for the investigation of physics around this scale.  It may be considered as an alternative way to prove the existence of a minimum length (for example $l_p\approx 1.6\cdot 10^{35} $meters required by string theory). The idea of a discrete structure of spacetime was first suggested by  Wheeler \cite{Wheeler} and well after by Wilson \cite{Wilson:1974sk}. The lattice formulation in a quantum field theory (QFT) is considered as a way of discretizing the path integral in order to make it well-defined. On a lattice, a  QFT becomes a quantum system
whose degrees of freedom consist of one field variable $\phi({\bf x_i})$
at each lattice point ${\bf x_i}\in \mathbb{Z}^d$. Thereby, the lattice is a way to regularize in the ultraviolet a QFT.  The  lattice structure does not violate  the quantum-mechanical structure of the
theory, but does not preserve the spacetime symmetries ( such as translations and rotations). The  numerical simulations of quantum field theories on Euclidean lattices have proven to be a very successful tool for studying nonperturbative phenomena. Consequently, a lot of effort has been put into the lattice formulation of quantum and field theories, see \cite{Amador:2016tec}-\cite{Gutzwiller:1981by}, \cite{Mehta:2009zv}-\cite{Cucchieri:2013nja}  and references therein. Hence, the discrete structure of spacetime is inherent in many models of quantum gravity, such as loop quantum gravity, noncommutative (NC) field theory, spin foam, black hole physics, random tensors models.

Recent results obtained in the framework of nonperturbative
string theory and quantum Hall effect, have boosted interest in a deeper
understanding of the role played by NC geometry
in different sectors of theoretical physics   \cite{Hellerman:2001rj}-\cite{Scholtz:2005vg}.  In
physics, the most important achievement of NC geometry was to overcome the distinction between continuous and discrete
spaces, in the same way that quantum mechanics unified the concepts of waves and particles.  However a  NC space is an intriguing  and revolutionary possibility that could have important consequences  in our conception of the quantum structure of nature. The description of 
 noncommutativity in quantum and  field theory can be
achieved  by replacing the ordinary product of functions in classical theory by the so called Moyal star product.  This can also be realized by defining the field theory on
a coordinates operators space that are intrinsically NC, which satisfy the commutation relation $[\hat  X^\rho,\hat X^\sigma]=i\theta^{\rho\sigma}$. The simplest case corresponds to where $\theta^{\rho\sigma}$
is a constant skew-symmetric matrix. In the present investigation we wish to define the noncommutativity of quantum theory, in which both coordinates and momenta are NC \cite{Snyder:1946qz}-\cite{Yang:1947ud}, i.e.  
\bea
[\hat X^\rho,\hat X^\sigma]&=&i\theta^{\rho\sigma},\cr
  [\hat X^\rho,\hat P^\sigma]&=&i\hbar_{eff}\delta^{\rho\sigma},\cr
 [\hat P^\rho,\hat P^\sigma]&=&i\bar\theta^{\rho\sigma}.
\eea 
The case $\theta^{\rho\sigma}=0=\bar\theta^{\rho\sigma}$ corresponds to ordinary quantum mechanics, for which $\hbar_{eff}=\hbar$ (the Planck constant). In the general  possible representations,  one obtained, from standard Bopp-shifts in the conventional canonical variables $\hat x^\rho$, $\hat p^\rho$, with nonvanishing commutators $[\hat x^\rho,\hat p^\sigma]=i\hbar\delta^{\rho\sigma}$,
$
\hat X^\rho= a\,\hat x^\rho+b\,\hat p^\rho,\,
\hat P^\rho= c\,\hat x^\rho+d\,\hat p^\rho,
$
where $a,b,c,d$ are constants.  With these transformations, all  Hamiltonians dynamics in NC space correspond to others problems in ordinary quantum space. As a motivating example, one could mention that the harmonic oscillator in NC space corresponds to the Landau problem in ordinary quantum space \cite{Gamboa:2001fg}-\cite{Dulat:2008eu}. It would therefore be interesting to investigate  the harmonic oscillator in the NC discrete space in which the continuous variables $\hat X^\rho$ and $\hat P^\rho$ become discrete with a spacing $\varepsilon$. It turns out that this question is not trivial, but may be solved in the perturbation to $\varepsilon$.  It is important to point out that the Landau problem on lattices has been extensively studied in the literature  see \cite{Valiente:2011zz}-\cite{Cucchieri:2013nja} and references therein. However, due to the infinite order derivatives which appear in the Schr\"odinger  equation, numerical solutions are the most developed. The very promizing analytic  approach proposed in \cite{Valiente:2011zz} proves to have several defects that we will explain in section \eqref{sec3}. In this paper the perturbative method is implemented to improve  these results. 

Let us recall very briefly known facts about the lattice oscillator in classical and quantum mechanics.
 Lattice oscillator systems are the standard model for the vibrational degrees of freedom,
known as phonons, in crystal lattices \cite{Drell:1977ge}-\cite{Gutzwiller:1981by}.  These phonons interact with the other degrees of
freedom, such as spins and electrons, in ways that often significantly modify their behavior.  The lattice  quantum theory is based on the non-differential calculus with discrete derivatives and integrals, which has been studied by several mathematicians and physicists \cite{Tarasov:2014iza}-\cite{Chernodub:2015ova},  and continues to be of interest for scientists nowadays. A few of its applications can be seen for instance in the study of non-local or time-dependent processes, as well as to model phenomena involving coarsegrained,  fractal spaces  and fractional systems as well as more simple systems such as harmonic oscillator \cite{He:2014cua}-\cite{Chernodub:2015ova}.  Most of the models of interacting quantum oscillators are related with solids such as ionic crystals containing localized
light particles oscillating in the field created by heavy ionic complexes.  The energy spectrum is  obtained by the ladder operators method, similar to the quantum harmonic oscillator problem. A lattice at a nonzero temperature has an energy that is not constant, but fluctuates randomly around some mean value. The thermodynamic properties and the quantum radiation maybe also examined closely (for a recent reviews see \cite{Jabbari:2012xi}-\cite{Jahan:2012wh}). Henceforth the study of the oscillator in the lattice, is a key to understand physics beyond continuous limit. We note that, several points of view have been developed and represented as the generalizations of the Heisenberg algebra to a discrete space. There are many lattice models which are reduced, classically to the same continuum theory in the zero lattice spacing, and this includes the $q$-deformations and those extensions \cite{Sun:1989tx}-\cite{Bhatia:2010qz}.

Our aim in the present work is to solve the quantum dynamics in the general NC discrete space and determine the eigenvalue problem of the corresponding oscillator Hamiltonian. The paper is organized as follows: In section \eqref{sec2} we briefly review some definitions and properties concerning the discrete differential calculus and its application to quantum mechanics. Next, we introduce the noncommutativity in this discrete space and show how the Heisenberg uncertainly relations are modified. Section \eqref{sec3} is devoted to the study of the $2d$ lattice harmonic oscillator in both commutative and NC quantum space. The corresponding eigenvalue problems are solved (i.e. in these two different cases).
In this section, we also deal with the thermodynamic
 behavior. In section \eqref{sec4} we make some remarks and conclude on our work. The direct computation of the states and energies of the oscillator in ordinary quantum space, performed using the Ladder operator method which appears in  \cite{Valiente:2011zz}, is also discussed.

\section{Discrete differential calculus and  lattice quantum mechanics}\label{sec2}

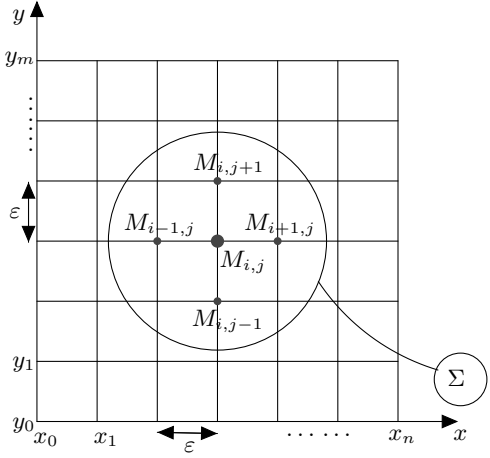
\begin{figure}\begin{center}
\definecolor{uuuuuu}{rgb}{0.27,0.27,0.27}
\begin{tikzpicture}[line cap=round,line join=round,>=triangle 45,x=0.8cm,y=0.8cm]
\draw [->] (0,-1) -- (0,6);
\draw [->] (0,-1) -- (7,-1);
\draw (0,5)-- (6,5);
\draw (6,5)-- (6,-1);
\draw (5,-1)-- (5,5);
\draw (6,4)-- (0,4);
\draw (0,3)-- (6,3);
\draw (1,5)-- (1,-1);
\draw (0,0)-- (6,0);
\draw (4,-1)-- (4,5);
\draw (0,2)-- (6,2);
\draw (0,1)-- (6,1);
\draw (3,-1)-- (3,5);
\draw (2,5)-- (2,-1);
\draw(3,2) circle (1.45cm);
\draw [->] (-0.14,1.98) -- (-0.14,2.98);
\draw [->] (-0.14,2.98) -- (-0.14,1.98);
\draw (-0.6,2.7) node[anchor=north west] {$\varepsilon$};
\draw (-0.55,6.0) node[anchor=north west] {$y$};
\draw (6.78,-1.0) node[anchor=north west] {$x$};
\draw (5.7,-1.0) node[anchor=north west] {$x_n$};
\draw (-0.68,5.30) node[anchor=north west] {$y_m$};
\draw (-0.20,-1.05) node[anchor=north west] {$x_0$};
\draw (-0.550,-0.8) node[anchor=north west] {$y_0$};
\draw (-0.55,0.22) node[anchor=north west] {$y_1$};
\draw (0.80,-1.05) node[anchor=north west] {$x_1$};
\draw (-0.3,4.8) node[anchor=north west] {\vdots};
\draw (-0.3,4.3) node[anchor=north west] {\vdots};
\draw (4.0,-1.05) node[anchor=north west] {\ldots};
\draw (4.6,-1.05) node[anchor=north west] {\ldots};
\draw (2.89,1.98) node[anchor=north west] {$M_{i,j}$};
\draw (2.45,3.60) node[anchor=north west] {$M_{i,j+1}$};
\draw (1.30,2.60) node[anchor=north west] {$M_{i-1,j}$};
\draw (2.43,1.00) node[anchor=north west] {$M_{i,j-1}$};
\draw (3.30,2.60) node[anchor=north west] {$M_{i+1,j}$};
\draw [shift={(7.72,3.36)}] plot[domain=3.73:4.42,variable=\t]({1*3.66*cos(\t r)+0*3.66*sin(\t r)},{0*3.66*cos(\t r)+1*3.66*sin(\t r)});
\draw [->] (2,-1.18) -- (3,-1.2);
\draw [->] (3,-1.2) -- (2,-1.18);
\draw (2.30,-1.20) node[anchor=north west] {$\varepsilon$};
\draw(7.04,-0.3) circle (0.35cm);
\draw (6.68,-0.0) node[anchor=north west] {$\Sigma$};
\begin{scriptsize}
\fill [color=uuuuuu] (3,2) circle (2.5pt);
\fill [color=uuuuuu] (3,3) circle (1.5pt);
\fill [color=uuuuuu] (3,1) circle (1.5pt);
\fill [color=uuuuuu] (2,2) circle (1.5pt);
\fill [color=uuuuuu] (4,2) circle (1.5pt);
\end{scriptsize}
\end{tikzpicture}
\caption{Two dimensions lattice and the  representation  of the subset $\Sigma$ with gives the interactions of the point $M_{i,j}$ with  his neighbours}\label{fig1}\end{center}
\end{figure}

In this section, we  review some basis properties  of the differential calculus on a $2d$ lattice (in particular we consider the case where $d=1$). It is based on the work in \cite{Jizba:2009qf}. For more details, one could also read \cite{Jizba:2011zz}-\cite{Jizba:2010zz} and the references therein. A lattice is a subset $\Gamma=([0,\ell]\times[0,\ell])\cap  \mathbb{Z}^2)$ of the plane $\mathbb{R}^2$ endowed with the discrete points $M_{n,m}:=M(x_n,y_m)$ such that the  coordinates $\{x_n\}_{n}$ and $\{y_m\}_{m}$, $n,m\in\mathbb{N}$ are  spacing by $\varepsilon<<1$ and $\varepsilon$ have the dimension of Planck lenght:  $[[\varepsilon]]\equiv [[l_p]]$.  We write $x_n=n\varepsilon$ and $y_m=m\varepsilon$, (see  figure \eqref{fig1}). 
We simplify the notation by setting $x_n:=x$ and $y_m:=y$.  Note that the discretization of space variables leads to a breaking of both
translational and rotational invariances, which are restored at the limit $\varepsilon\rightarrow 0$. It is also possible to define the discrete  translation  as $x'_n=x_n\pm k\varepsilon,\,y_n'=y_n\pm k\varepsilon,\,\, k\in\mathbb{Z}$ and such that the translation symmetry on the lattice is preserved. The same procedure can be done in the case of space rotation. In conclusion, instead of continuous
Lorentz, translation and rotation symmetry we have the discrete symmetries of the lattice. This need not be a problem, since we can recover the continuous symmetries at low energies.

The most obvious choice to discretize the continuous derivatives is to use the discrete symmetric derivative:  Naturally, the derivatives $\partial_x:=\frac{\partial}{\partial x}$ and $\partial_y:=\frac{\partial}{\partial y}$ are replaced  by
the forward and backward difference operators  $\ds_j^+$ and $\mathbb d_j^-$, $j=x,y$, also called the left and right non-differential operators acting on the two variables dependent function $f(x,y)$ as
\bea
\ds_{x}^{+}f(x,y)&=&\frac{1}{\varepsilon}\Big[f(x + \varepsilon, y) - f(x, y)\Big],\\
\ds_{x}^{-}f(x,y)&=&\frac{1}{\varepsilon}\Big[ f(x, y) - f(x - \varepsilon, y)\Big], \\
\ds_{y}^{+}f(x,y)&=&\frac{1}{\varepsilon}\Big[f(x,y + \varepsilon) - f(x, y)\Big],\\
\ds_{y}^{-}f(x,y)&=&\frac{1}{\varepsilon}\Big[ f(x, y) - f(x,y - \varepsilon)\Big].
\eea
Observe that, at the limit $\varepsilon\rightarrow 0$, $\ds_{x}^{+}=\ds_{x}^{-}=\partial_x$ and $\ds_{y}^{+}=\ds_{y}^{-}=\partial_y$. All the computations performed here can be generalized to arbitrary dimensions $d>1$. The operators $\ds_{x}^{+},\,\,\ds_{x}^{-}$, $\ds_{y}^{+},\,\,\ds_{y}^{-}$  are related to the translation operators  in the $x$ and $y$ directions  denoted by $\tau^\varepsilon_x$ and $\tau^\varepsilon_y$ with group parameter $\varepsilon$ as:
\bea
\tau^\varepsilon_x=e^{\varepsilon \partial_x},\quad \tau^\varepsilon_y=e^{\varepsilon \partial_y},
\eea
and such that
 $\tau^\varepsilon_x f(x,y)=f(x+\varepsilon,y)$ and $\tau^\varepsilon_y f(x,y)=f(x,y+\varepsilon)$. We get the followings identities, for $ j=x,y$:
\bea
\ds_{j}^{+}  &=& \frac{1}{\varepsilon}\big( \tau^\varepsilon_j - \mathbb{1}\big),\quad 
\ds_{j}^{-} = -\frac{1}{\varepsilon}\big( \tau^{-\varepsilon}_j - \mathbb{1}\big).
\eea
The following generalized Leibnitz rules hold: 
\begin{eqnarray}
\ds_{j}^{+} (f g) &=& \frac{1}{\varepsilon}\big( \tau^\varepsilon_j f  \tau^\varepsilon_j  g - f g\big)\cr
&=&g\ds_{j}^{+}f+f\ds_{j}^{+} g+\varepsilon \ds_{j}^{+} f \ds_{j}^{+}g,\\
\ds_{j}^{-} (f g) &=& \frac{1}{\varepsilon}\Big( fg-\tau^{-\varepsilon}_j f  \tau^{-\varepsilon}_j  g \Big)\cr
&=&g\ds_{j}^{-}f+f\ds_{j}^{-} g-\varepsilon \ds_{j}^{-} f \ds_{j}^{-}g,
\end{eqnarray}
and they are reduced to usual Leibnitz rules as $\varepsilon\rightarrow 0$.  One can also define the discrete Laplacian as 
\bea\label{taka}
\bm{\ds}^2 &=& \ds_{x}^{+} \ds_{x}^{-} + \ds_{y}^{+} \ds_{y}^{-}\cr
&=&\frac{2}{\varepsilon^2}\Big[\cosh({\varepsilon \partial_x}) + \cosh({\varepsilon \partial_y}) - 2\Big].
\eea 
 This quantity  plays an important role when defining the kinetic part of the Hamiltonian both in the classical and quantum description of dynamic systems on the $2d$   lattice i.e.  $\hat H=-\Omega \bm{\ds}^2+\hat V(x,y) $, $\Omega\in\mathbb{R}$ where $\hat V(x,y)$ is the interaction potential.

We are now ready to define the relation between non differential geometry and quantum mechanics on the lattice (for more detail see \cite{Jizba:2011zz}-\cite{Jizba:2009qf}). Let us consider the Hilbert space $\mathcal{H}$ endowed with the Hermitian structure
\begin{eqnarray}\label{Hermitian}
\langle f|g\rangle = \varepsilon^2 \sum_{x,y} f^{*}(x,y) g(x,y).
\end{eqnarray}
Consider the subset $\mathcal{H}_{\otimes}$ of $\mathcal H$ in which the states $f(x,y)$ can be decomposed into $f_1(x)\otimes f_2(y)$. Then  $\mathcal{H}_\otimes=\mathcal{H}_x\otimes \mathcal{H}_y\equiv L^2(\mathbb{R},dx)\bigotimes L^2(\mathbb{R},dx)$ and
the tensors product operators 
$\bm\ds^+ = \ds_x^+\otimes \ds_y^+$ and $\bm\ds^- = \ds_x^-\otimes \ds_y^-$ acting on  $\mathcal{H}_\otimes$ are  not Hermitian. We find that $\Big(i\bm\ds^+\Big)^\dagger = i \bm\ds^-$.
 However the operators  $\bm\ds^- \bm\ds^+ = \bm\ds^+\bm \ds^-$, which corresponds to the fact that the laplacian \eqref{taka} is Hermitian. Now let us consider the positions and momenta operators $\hat x^\varepsilon$ and $\hat y^\varepsilon$,  defined by the eigen-equations
\begin{eqnarray}\label{socol}
\big(\hat x^\varepsilon f\big)(x,y) = x f(x, y), \,\,\,\,
\big(\hat y^\varepsilon f\big)(x,y) = y f(x, y)
\end{eqnarray}
 and the momentum  $\hat p_x^\varepsilon$ and $\hat p_y^\varepsilon$  as 
\begin{eqnarray}\label{momentx}
\Big(\hat p_{x}^\varepsilon f \Big)(x,y) &=& -\frac{i \hbar}{2}\Big(\ds_x^+  + \ds_x^-  \Big)f(x,y)\cr
&=&-\frac{i \hbar}{\varepsilon}\sinh(\varepsilon \partial_x ) f(x,y), 
 \\
\label{momenty}\Big(\hat p_{y}^\varepsilon f \Big)(x,y) &=& -\frac{i \hbar}{2}\Big(\ds_y^+ + \ds_y^- \Big)f (x,y) \cr
&=& -\frac{i \hbar}{\varepsilon}\sinh(\varepsilon \partial_y ) f(x,y).
\end{eqnarray}
The operators \eqref{momentx} and \eqref{momenty} are  Hermitian and  have the nonvanish commutation relations:
\begin{eqnarray}\label{trz}
\Big[\hat x^\varepsilon, \hat p_x^\varepsilon\Big] = i \hbar\cosh\big(\varepsilon \partial_x\big),\,\,\,\,
\Big[\hat y^\varepsilon, \hat p_y^\varepsilon\Big] = i \hbar\cosh\big(\varepsilon \partial_y\big). 
\end{eqnarray}
 Observe that in the equations \eqref{momentx}, \eqref{momenty} and \eqref{trz} we have used the following definitions:
\bea
&\cosh(\varepsilon\partial_j)=\frac{e^{\varepsilon\partial_j}+e^{-\varepsilon\partial_j}}{2},\, \sinh(\varepsilon\partial_j):=\frac{e^{\varepsilon\partial_j}-e^{-\varepsilon\partial_j}}{2},\cr
&j=x,y.\nonumber
\eea
Using the Taylor expansion of these two operators functions the commutation relations \eqref{trz}, which are a consequence of the definitions  \eqref{socol}, \eqref{momentx} and \eqref{momenty}, 
 are nothing but the sum of polynomial operators depending on the ordinary momenta $\hat p_x = - i \hbar \partial_x$ and $\hat p_y = - i \hbar \partial_y$, such that the limit as $\varepsilon\to 0$ is well given as canonical commutation relations between coordinates and momenta (see relations \eqref{sarmat1} for more details.)

As a quantum system the space of states  of a physical model defined with the  Hamiltonian $H=\frac{|\hat p^{\varepsilon}|^2}{2m}+V(|\hat x^{\varepsilon}|)$ should then provide a linear representation space of the generalized Heisenberg algebra, equipped with the Hermitian inner product \eqref{Hermitian} for which these two operators be self-adjoint.
The uncertainly relation  is now generalized as
\begin{eqnarray}\label{bien1}
\Delta \hat x^\varepsilon \Delta \hat p_x^\varepsilon \ge \frac{\hbar}{2}\Big|\sum_{n=0}^{\infty} \frac{(-1)^n}{(2n)!}\Big(\frac{\varepsilon}{\hbar}\Big)^{2n}\langle \hat p_x^{2n}\rangle\Big|,\\ \Delta \hat y^\varepsilon \Delta \hat p_y^\varepsilon \ge \frac{\hbar}{2}\Big|\sum_{n=0}^{\infty} \frac{(-1)^n}{(2n)!}\Big(\frac{\varepsilon}{\hbar}\Big)^{2n}\langle \hat p_y^{2n}\rangle\Big|
\end{eqnarray}
where  $\hat p_x = - i \hbar \partial_x$, $\hat p_y = - i \hbar \partial_y$ correspond to the momentum operators in ordinary quantum mechanics,
the first order expansion of the relations \eqref{bien1} gives
\begin{eqnarray}\label{un1}
\Delta \hat x^\varepsilon \Delta \hat p_x^\varepsilon \ge = \frac{\hbar}{2}\Big(1+\frac{\varepsilon^{2}}{2\hbar^2}|\langle \hat p_x^{2}\rangle|+\mathcal O(\varepsilon^2)\Big),\\\label{un2}
\Delta \hat y^\varepsilon \Delta \hat p_y^\varepsilon \ge = \frac{\hbar}{2}\Big(1+\frac{\varepsilon^{2}}{2\hbar^2}|\langle \hat p_y^{2}\rangle|+\mathcal O(\varepsilon^2)\Big).  
\end{eqnarray}
 The  relations \eqref{un1} and \eqref{un2} correspond to the uncertainty relations predicted by string theory as a correction of the usual uncertainty relations between coordinates and momenta.  It is probably one of the greatest interest in the study of the minimum lenght quantum theory  \cite{Kempf:1998gk}-\cite{Kempf:1996fz},  (see also \cite{Doplicher:1994tu}-\cite{Gross:1987ar}).  This implies the following commutation relation 
\bea\label{sarmat1}
&&[ \hat x^\varepsilon, \hat p_x^\varepsilon]=i\hbar\Big(1+\frac{\varepsilon^{2}}{2\hbar^2}(\hat p_x^\varepsilon)^2+\mathcal O(\varepsilon^2)\Big),\cr
&&[ \hat y^\varepsilon, \hat p_y^\varepsilon]=i\hbar\Big(1+\frac{\varepsilon^{2}}{2\hbar^2}(\hat p_y^\varepsilon)^2+\mathcal O(\varepsilon^2)\Big).
\eea
The parameter $\varepsilon$ is chosen such that the corresponding uncertainty relations \eqref{un1} and \eqref{un2} imply a finite minimal uncertainty $\Delta
\hat x^\varepsilon_0>0$ and $\Delta \hat y^\varepsilon_0>0$.

Recall that we are interested in investigating the behavior of the oscillator model in noncommutative space. For this purpose, we have to determine the energy spectrum of the Hamiltonian for the small value of the spacing $\varepsilon$.  We use the ``{\it capital }'' notation to specify the NC quantum operators, such as coordinates and momenta. Suppose that the NC variables are related to the commutative coordinates operators by the relations:
\begin{eqnarray}
 \hat{X}^\varepsilon = \hat x^\varepsilon - \frac{\theta}{2\hbar} \hat p_y^\varepsilon ,\quad
\hat{Y}^\varepsilon = \hat y^\varepsilon + \frac{\theta}{2\hbar} \hat p_x^\varepsilon \\
 \hat{P}_x^\varepsilon = \hat p_x^\varepsilon + \frac{\bar{\theta}}{2\hbar} \hat y^\varepsilon,\quad
 \hat{P}_y^\varepsilon = \hat p_y^\varepsilon - \frac{\bar{\theta}}{2\hbar} \hat x^\varepsilon
\end{eqnarray}
The commutation relations between coordinates and momenta are then taken to be:
\begin{eqnarray}\label{ssss}
\Big[\hat{X}^\varepsilon, \hat{Y}^\varepsilon\Big] = i \theta \hat{I}_0^\varepsilon,\,\, 
\Big[\hat{P}_x^\varepsilon, \hat{P}_y^\varepsilon\Big] =  i \bar{\theta}  \hat{I}_0^\varepsilon,\cr
 \Big[\hat{X}^\varepsilon, \hat{P}_x^\varepsilon\Big] =  i \hbar \hat{I}_1^\varepsilon,\,\,
 \Big[\hat{Y}^\varepsilon, \hat{P}_y^\varepsilon\Big] =  i \hbar \hat{I}_2^\varepsilon,
\end{eqnarray}
where the  operators $\hat{I}_0^\varepsilon ,   \hat{I}_1^\varepsilon, \hat{I}_2^\varepsilon$ are given by
\bea
\hat{I}_0^\varepsilon =  \frac{1}{2} \Big[\cosh\big(\varepsilon \partial_x\big) + \cosh\big(\varepsilon \partial_y\big)\Big], \\ \hat{I}_1^\varepsilon = \cosh\big(\varepsilon \partial_x\big)  + \frac{\theta \bar{\theta}}{4\hbar^2} \cosh\big(\varepsilon \partial_y\big), \\ \hat{I}_2^\varepsilon = \cosh\big(\varepsilon \partial_y\big) + \frac{\theta \bar{\theta}}{4\hbar^2} \cosh\big(\varepsilon \partial_x\big).
\eea
The followings uncertainly relations are well satisfied:
\begin{eqnarray}
\Delta \hat{X}^\varepsilon \Delta \hat{Y}^\varepsilon \ge  \frac{\theta}{4}\Big|\sum_{n=0}^{\infty} \frac{\varepsilon^{2n}}{(2n)!} \Big(\langle\partial_x^{2n} \rangle +  \langle \partial_y^{2n} \rangle\Big)\Big|,\\
\Delta \hat{P}_x^\varepsilon \Delta \hat{P}_y^\varepsilon \ge  \frac{\bar\theta}{4}\Big|\sum_{n=0}^{\infty} \frac{\varepsilon^{2n}}{(2n)!} \Big(\langle\partial_x^{2n} \rangle +  \langle \partial_y^{2n} \rangle\Big)\Big| 
\end{eqnarray}
\begin{eqnarray}
\Delta \hat{X}^\varepsilon \Delta \hat{P}_x^\varepsilon  \ge \frac{\hbar}{2}\Big|\sum_{n=0}^{\infty} \frac{\varepsilon^{2n}}{(2n)!} \Big(\langle \partial_x^{2n} \rangle + \frac{\theta \bar{\theta}}{4\hbar^2} \langle \partial_y^{2n} \rangle \Big)\Big|
,\\
\Delta \hat{Y}^\varepsilon \Delta \hat{P}_y^\varepsilon  \ge \frac{\hbar}{2}\Big|\sum_{n=0}^{\infty} \frac{\varepsilon^{2n}}{(2n)!}\Big(\frac{\theta \bar{\theta}}{4\hbar^2} \langle \partial_x^{2n} \rangle + \langle \partial_y^{2n} \rangle \Big) \Big|
\end{eqnarray}
The first order perturbation gives:
\bea
\Delta \hat{X}^\varepsilon \Delta \hat{Y}^\varepsilon \ge\frac{\theta}{4}\Big[1-\frac{\varepsilon^2}{2\hbar^2}(p_x^2+p_y^2)\Big],\\ \Delta \hat{P}_x^\varepsilon \Delta \hat{P}_y^\varepsilon \ge  \frac{\bar\theta}{4}\Big[1-\frac{\varepsilon^2}{2\hbar^2}(p_x^2+p_y^2)\Big]
\eea
\bea
\Delta \hat{X}^\varepsilon \Delta \hat{P}_x^\varepsilon  \ge \frac{\hbar}{2}\Big[1+\frac{\theta \bar{\theta}}{4\hbar^2}-\frac{\varepsilon^2}{2\hbar^2}\Big(p_x^2+\frac{\theta \bar{\theta}}{4\hbar^2}p_y^2\Big)\Big],\\ \Delta \hat{Y}^\varepsilon \Delta \hat{P}_y^\varepsilon  \ge \frac{\hbar}{2}\Big[1+\frac{\theta \bar{\theta}}{4\hbar^2}-\frac{\varepsilon^2}{2\hbar^2}\Big(p_y^2+\frac{\theta \bar{\theta}}{4\hbar^2}p_x^2\Big)\Big]
\eea
which corresponds to the generalization of the uncertainty relation for physics at the Planck scale predicted by the string theory and given in \eqref{un1} and \eqref{un2}.  Let us mention that for a sufficiently small constant $\frac{\varepsilon^2}{2\hbar^2}$, the correction term in the uncertainty relations are negligible at present day experimentally accessible scales. By choosing this parameter appropriately we obtain a cut-off at the string or at the Planck scales. This type of ultraviolet
cut-off was introduced into quantum field theory in \cite{Kempf:1996ss} and then into inflationary cosmology in \cite{Kempf:2000ac}.  As readily checked, this implies a minimal uncertainly in the positions $\Delta \hat X^\varepsilon$ and $\Delta \hat Y^\varepsilon$ namely  $\Delta \hat X_0^\varepsilon$ and $\Delta \hat Y_0^\varepsilon$ which are given by
\bea
\Delta \hat X_0^\varepsilon= \Delta \hat Y_0^\varepsilon\propto\varepsilon\equiv l_p.
\eea
 However, we have a maximal dispersion for the momenta
\bea
\Delta \hat P_{x0}^\varepsilon= \Delta \hat P_{y0}^\varepsilon\propto \frac{1}{\varepsilon\sqrt{\varepsilon}},
\eea
which is infinite in the continuum limit. One can also remark that the noncommutativity of the coordinates operators $\hat X^\varepsilon$ and $\hat Y^\varepsilon$ will not be necessary for the appearance of a finite minimal uncertainty $\Delta \hat X_0^\varepsilon$ and $\Delta \hat Y_0^\varepsilon$:
See \cite{Bojowald:2011jd}-\cite{Kempf:1996nm} for more details.

\section{Perturbation method for the harmonic oscillator on a lattice }\label{sec3}
In this section, the low energy approximation is given for the harmonic oscillator. First 
we examinate the case of ordinary quantum space defined with the commutation relation \eqref{trz}. The next part is devoted to the same computation where we have to consider  noncommutativity in general case given in \eqref{ssss}.
\subsection{Harmonic oscillator in the ordinary quantum space lattice}
 Consider the subset $\Sigma$ of the lattice given in figure \eqref{fig1}, in which the point $M_{k,j}$ interact with the four  neighbours $M_{k+1,j},\,M_{k-1,j},\, M_{k,j+1},\, M_{k,j-1}$. These interactions are supposed to be  harmonic and the Hamiltonian becomes
\beq
\hat H_{\varepsilon,kj} = \frac{1}{2 m}\Big[\big(\hat p_{x_k}^{\varepsilon}\big)^2 + \big(\hat p_{y_j}^{\varepsilon}\big)^2\Big] + \frac{m \omega^2 }{2}\Big[\big(\hat x_k^{\varepsilon}\big)^2 +\big(\hat y_j^{\varepsilon}\big)^2\Big].
\eeq
The total Hamiltonian that describes the oscillation of all points of the lattice is
\bea\label{hamil}
\hat H_\varepsilon=\sum_{\{k,j\}} \hat H_{\varepsilon,kj}\in L(\mathcal{H}_\otimes),
\eea
where $L(\mathcal{H}_\otimes)$ is the set of linear operators on the Hilbert space $\mathcal{H}_\otimes$. For simplicity, the sum in expression \eqref{hamil} will not be written.  Then, using \eqref{momentx} and \eqref{momenty}, expression \eqref{hamil}  becomes
\begin{eqnarray}\label{hamilnew}
\hat H_\varepsilon &=& -\frac{\hbar^2}{2 m \varepsilon^2} \Big[\sinh^2(\varepsilon \partial_x ) + \sinh^2(\varepsilon \partial_y )\Big] \cr
&&+ \frac{m \omega^2 }{2}\Big[(\hat{x}^{\varepsilon})^2 + (\hat{y}^{\varepsilon})^2\Big]
\end{eqnarray}
Let us turn now to the solution of the eigenvalues problem by using the corresponding partial differential equation, which is explicitly given in \cite{Valiente:2011zz}:
\bea\label{widelu}
\hat H_\varepsilon\phi_n(x,y)=E_n\phi_n(x,y).
\eea
Now, we consider the wave function in the Fourier space. The coordinates and momenta operators are given by
\bea
&&\hat{p}_x^{\varepsilon} = \frac{\hbar}{\varepsilon} \sin(\varepsilon k_x), \quad \hat{p}_y^{\varepsilon} = \frac{\hbar}{\varepsilon} \sin(\varepsilon k_y), \\ 
&&\hat{x}^{\varepsilon} = i \frac{\partial}{\partial k_x}, \quad \hat{y}^{\varepsilon} = i \frac{\partial}{\partial k_y}
\eea
where  $k_x, k_y$, taken on the Brillouin zone $ ]-\frac{\pi}{\varepsilon}, \frac{\pi}{\varepsilon}]$ are called the quasi-momenta.

\subsubsection{Discussion about the difficulties to provide algebraic solution}
In this subsection we give in detail  the set of difficulties that comes in trying to determine the spectrum of the Hamiltonian \eqref{hamilnew} or in the search of the solution of \eqref{widelu}. It is based on the work given in reference \cite{Valiente:2011zz} from which the author provide one alternative way  to solve \eqref{widelu}. We will  show the non-consistency of this method and propose to use perturbative solution.
Let us define the lattice analogue of the harmonic oscillator ``annihilation   and creation'' operators as:
\begin{eqnarray}
\label{ha1}&& \hat{a}_x = \frac{1}{\sqrt{2\hbar m \omega}} (m\omega \hat{x}^\varepsilon + i\hat{p}_x^\varepsilon), \\
&& \hat{a}_x^\dag = \frac{1}{\sqrt{2\hbar m \omega}} (m\omega \hat{x}^\varepsilon - i\hat{p}_x^\varepsilon) \\
&&  \hat{a}_y = \frac{1}{\sqrt{2\hbar m \omega}} (m\omega \hat{y}^\varepsilon + i\hat{p}_y^\varepsilon), \\
\label{ha4}&& \hat{a}_y^\dag = \frac{1}{\sqrt{2\hbar m \omega}} (m\omega \hat{y}^\varepsilon - i\hat{p}_y^\varepsilon),
\end{eqnarray}
 which are supposed to diagonalize the Hamiltonian $H_\varepsilon=H_{x,\varepsilon}+H_{y,\varepsilon}$ only in the continuous limit $\varepsilon\to 0$, with
\begin{eqnarray}\label{Ham1}
H_{x,\varepsilon} = \frac{(\hat{p}_x^{\varepsilon})^2}{2 m}  + \frac{m \omega^2  (\hat{x}^{\varepsilon})^2}{2},\\
 \label{Ham2} H_{y,\varepsilon}= \frac{(\hat{p}_y^{\varepsilon})^2}{2 m}  + \frac{m \omega^2  (\hat{y}^{\varepsilon})^2}{2}.
\end{eqnarray}

 Before we start our discussion to elucidate the problems that arise  when we want to solve the eigenvalue problem of the Hamiltonian    \eqref{Ham1} and \eqref{Ham2}, and why we need to provide new method, let us remark that the so called annihilation and creation operators given in \eqref{ha1}-\eqref{ha4} do not satisfy the usual canonical commutation relation due to the presence of the lattice spacing $\varepsilon$. We get  $[\hat{a}_x,\hat{a}_x^\dag]=\cos(k_x\varepsilon)$ and $[\hat{a}_y,\hat{a}_y^\dag]=\cos(k_y\varepsilon)$, such that the limit $\varepsilon\to 0$ leads to $[\hat{a}_x,\hat{a}_x^\dag]_{\varepsilon\to 0}=\mathbb{1}$, $[\hat{a}_y,\hat{a}_y^\dag]_{\varepsilon\to 0}=\mathbb{1}$. The operators $\hat{a}_x,\hat{a}_x^\dag$ and $\hat{a}_y,\hat{a}_y^\dag$ are therefore not the Ladder operators.  Now let us use the following approximation $\varepsilon\to 0$,  and from which we can consider the operators \eqref{ha1}-\eqref{ha4} as the Ladder operators.  Let $\phi_0=\phi_{0x}\otimes\phi_{0y}$ be the fundamental eigen-state such that
$
\hat{a}_x \otimes \hat{a}_y (\phi_{0x}\otimes\phi_{0y})=\hat{a}_x \phi_{0x}\otimes\hat{a}_y \phi_{0y}=0.
$
It leads  to the solution
\bea
\phi_{0x} = C_x \exp\Big[\frac{\hbar}{m\omega \varepsilon^2} \cos(\varepsilon k_x)\Big],\\
\phi_{0y} = C_y \exp\Big[\frac{\hbar}{m\omega \varepsilon^2} \cos(\varepsilon k_y)\Big],
\eea
where the constants $C_x$ and $C_y$, are given, using the normalization condition
\bea
\int_{-\pi/\varepsilon}^{+\pi/\varepsilon} d k_x\, \phi_{0x}^2=\int_{-\pi/\varepsilon}^{+\pi/\varepsilon} d k_y \,\phi_{0y}^2=1,\cr
 C_x =C_y=\Big[2 \pi J_0 \Big( \frac{2 i\hbar}{m\omega \varepsilon^2}\Big)\Big]^{-\frac{1}{2}},
\eea
such that
\bea\label{stateini}
\phi_{0}(k_x,k_y) &=&\Big[2 \pi J_0 \Big(\frac{2 i \hbar}{m\omega \varepsilon^2}\Big)\Big]^{-1} \exp\Big[\frac{\hbar}{m\omega \varepsilon^2} \cos(\varepsilon k_x)\Big]\cr
&&\otimes \exp\Big[\frac{\hbar}{m\omega \varepsilon^2} \cos(\varepsilon k_y)\Big]
\eea
where $J_0$ stands for the first kind Bessel function.   Remark that in our  solution \eqref{stateini}, the limit $\varepsilon\to 0$ is not well defined (i.e. $\lim_{\varepsilon\to 0}\phi_{0}(k_x,k_y)=\infty$). The same pathology appears in the reference \cite{Valiente:2011zz} after computing the normalization constant.
Replacing the solution \eqref{stateini} in the eigenvalue equation \eqref{widelu}, the fundamental energy becomes
\bea
E_0(k_x,k_y)=\frac{\hbar\omega}{2}(\cos(\varepsilon k_x)+\cos(\varepsilon k_y)).
\eea 
The others states may be determined order by order using the creation operators $\hat{a}_x\otimes \hat{a}_y$. Let us now comment this result. First of all, recall that the limit $\varepsilon \rightarrow 0$ is not well defined using \eqref{stateini}. Also, the energy $E_0(k_x,k_y)$ depends on $k_x$ and $k_y$, which means that after the inverse Fourier transformation we get
\bea
E_0(x,y)&=&\frac{1}{(2\pi)^2}\Big|\int_{-\frac{\pi}{\varepsilon}}^{\frac{\pi}{\varepsilon}}e^{i (k_x x+k_yy)}E_0(k_x,k_y) dk_x dk_y\Big|\cr
&=&\frac{\hbar\omega}{2\pi^2}\Big|\kappa(x,y)+\kappa(y,x)\Big|,
\eea
where
$$
\kappa(x,y)=\frac{x\sin(\pi x/\varepsilon)\sin(\pi y/\varepsilon)}{y(\varepsilon^2-x^2)},
$$
 which depends on the coordinates functions $x$ and $y$. It should also be noted that the continuous limit is given by
\beq
\lim_{\varepsilon\rightarrow 0}E_0(x,y)=\hbar\omega\delta(x)\delta(y),
\eeq
which is also not well defined at the ground state energy of the harmonic oscillator, owing to the presence of the Dirac delta function.  
 All These pathology are a consequence of the treatment we have done  with  the  definition of lattice analogue of the annihilation and creation operators \eqref{ha1}-\eqref{ha4}. In attempting to fill these gaps, we'll consider the perturbation method to derived the eigenvalue equation \eqref{widelu}.

\subsubsection{Perturbation method and solution}
Considering the  Taylor expansion of $\sinh^2(\varepsilon \partial_x )+\sinh^2(\varepsilon \partial_y) $,
the first order expansion to $\varepsilon^2$ of the Hamiltonian $H_\varepsilon$ takes the form
\begin{eqnarray}\label{samdino31}
\hat H_\varepsilon &=& -\frac{\hbar^2}{2 m} \Big(\partial_x^2 + \partial_y^2\Big) + \frac{m \omega^2 }{2}\Big[\big(\hat x^{\varepsilon}\big)^2 +\big(\hat y^{\varepsilon}\big)^2\Big]\cr
&&- \frac{\varepsilon^2 \hbar^2}{6 m}  \Big(\partial_x^4 + \partial_y^4\Big) + \mathcal{O}(\varepsilon^2)\cr
&=&\hat H_0 + \varepsilon^2  \hat W+ \mathcal{O}(\varepsilon^2).
\end{eqnarray}
 $\hat H_0$ corresponds to the harmonic oscillator Hamiltonian in ordinary space and 
 $\hat W$  is considered to be the perturbation term. Thus, we can introduce the new annihilation and creation operators defined in the limit $\varepsilon=0$ as:
\begin{eqnarray}
 &&\hat b_x = \frac{1}{\sqrt{2\hbar m \omega}}\Big(i\hat p_x + m\omega \hat x^\varepsilon\Big)=\lim_{\varepsilon\to 0}\hat a_x,\\
  &&\hat b^\dag_x = \frac{1}{\sqrt{2\hbar m \omega}}\Big(- i\hat p_x + m\omega \hat x^\varepsilon\Big)=\lim_{\varepsilon\to 0}\hat a_x^{\dag}, \\
&&  \hat b_y = \frac{1}{\sqrt{2\hbar m \omega}}\Big(i\hat p_y + m\omega \hat y^\varepsilon\Big)=\lim_{\varepsilon\to 0}\hat a_y,\\
 &&\hat b^\dag_y = \frac{1}{\sqrt{2\hbar m \omega}}\Big(- i\hat p_y + m\omega \hat y^\varepsilon\Big)=\lim_{\varepsilon\to 0}\hat a_y^\dag,
\end{eqnarray}

such that the canonical commutation relation   $\big[\hat b_i, \hat b_j^\dag\big] = \mathbb{1}\delta_{ij}$ is well  satisfied, and in these new coordinates  $\hat H_0$ and $\hat W$ take the form 
\bea
\hat  H_0 &=& \hbar \omega \Big(\hat b_x \hat b_x^\dag + \hat b_y \hat b_y^\dag - 1\Big),
\\
\hat W &=& - \frac{m \omega^2}{24} \Big((\hat b_x - \hat b_x^\dag)^4 + (\hat b_y - \hat b_y^\dag)^4\Big).
\eea
One constructs the Fock states as   $\{\ket{n_x,n_y;0}=\ket{n_x}\otimes\ket{n_y}\in\mathfrak{H},\, n_x,n_y\in\mathbb{N}\}$ such that the followings relations hold:
\beq
\hat b_x\otimes\hat b_y \ket{n_x,n_y;0}=\sqrt{n_x n_y}\ket{n_x-1,n_y-1;0},
\eeq
\beq
\hat b_x^\dag\otimes\hat b_y^\dag \ket{n_x,n_y;0}=\sqrt{(n_x+1) (n_y+1)}\ket{n_x+1,n_y+1;0}.
\eeq
and such that $\hat b_x\otimes\hat b_y \ket{0,0;0}=0$.
The  eigen-equation 
 $\hat H_0 \ket{n_x, n_y;0} = E_{n_x, n_y}^0 \ket{n_x, n_y;0}$, gives  $n_x+n_y+1$ degenerate states such that
$
E_{n_x,n_y}^0 = \hbar \omega (n_x + n_y + 1). 
$ 
We denote these degenerate states  as  $\ket{n_x,n_y;0}^j$, such that for $n=n_x+n_y$:
\beq\label{eqeq2}
\ket{n_x, n_y;0}^j= \Big\{\ket{n-j,j;0}; \,j =0,1,2,\cdots, n\Big\}.
\eeq

Consider now the vector
$
\ket{N} = \sum_{j=0}^{n} c_j \ket{n_x, n_y;0}^j + \varepsilon^2 \ket{n_x, n_y;1} + \mathcal O(\varepsilon^2)$  and the energy $E_{\varepsilon,n_x,n_y} = E_{n_x,n_y}^0 + \varepsilon^2 E_{n_x,n_y}^1 + \mathcal O(\varepsilon^2)$,
which solve the eigen-problem  $\hat H_{\varepsilon} \ket{N} = E_{\varepsilon,n_x,n_y} \ket{N}$, 
 and $c_j\in\mathbb{C},j=0,1,\cdots, n$. We have  the orthogonality relation 
$\sum_{j=0}^n c_j \; ^k\braket{0;n_x, n_y}{n_x, n_y;0}^j = \delta^{jk}$ and  the first order correction of the energy  i.e. $E_{n_x,n_y}^1$ is determined by the following linear homogeneous system
\begin{eqnarray}\label{quoisa}
\sum_{j=0}^n c_j \; {}^k\bra{0;n_x, n_y} \hat W \ket{n_x, n_y;0}^j = E_{n_x,n_y}^1 c_k.
\end{eqnarray}
While the above system is completely determine by the matrix $G$ such that $\det{G} = 0$
\begin{eqnarray}\label{Gmat}
G : \left\{\begin{array}{cc}
&G_{kk} = {}^k\bra{0;n_x, n_y} \hat W \ket{n_x, n_y;0}^k -  E_{n_x,n_y}^1 \\
&G_{kj} = {}^k\bra{0;n_x, n_y} \hat W \ket{n_x, n_y;0}^j \quad \textrm{for} \quad k \neq j
\end{array} \right.
\end{eqnarray}
 A few computation shows that $G$ is diagonal matrix and the diagonal elements  $G_{kk}$ are given by
\beq
G_{kk}=-\frac{m \omega^2}{4}\Big(n(n+1) - 2k n + 2k^2 + 1\Big) -  E_{n}^1
\eeq
where $E_n^1=E_{n_x,n_y}^1$, satisfies the following equation
\beq
 \prod_{k=0}^n \Big[-\frac{m \omega^2}{4} \Big(n^2 + n(1 - 2 k) + 2 k^2 + 1 \Big) - E_{n}^{1k}\Big] = 0.
\eeq
The index ``$k$'' in $E _{n}^{1 k}$ is used to specify the  degeneracy, such that
\bea
E_{\varepsilon,n}^{k} &=& \hbar \omega \Big(n + 1\Big) 
- \frac{m \omega^2 \varepsilon^2}{4} \Big[n^2 + n(1 - 2 k) + 2 k^2 + 1 \Big] \cr
&&+ \mathcal{O}(\varepsilon^2).
\eea
$\ket{n_x, n_y;1}$ can be computed in the same manner. Let  $ \mathcal{D} = \{n- j;\; 0 \le j \le n\} $ and $\mathcal{D'} = \{j ;\;  0 \le j \le n\}$, we get 
\beq
\ket{n_x, n_y;1} = - \sum_{j=0}^{n}\sum_{\substack{m_x, m_y \\ m_x \notin \mathcal{D} \\ m_y \notin \mathcal{D'}}}   \frac{ c_j\langle \hat W\rangle^j_{mn}}{E_{m_x,m_y}^0 - E_{n_x,n_y}^0} \ket{m_x,m_y;0}=0.
\eeq
where 
$\langle \hat W\rangle^j_{mn}=\bra{0;m_x,m_y}\hat W\ket{n_x, n_y;0}^j$.

\begin{widetext}

Let us  now deal with the thermodynamic behavior of the oscillator model. 
 First let us recall that the thermodynamic behavior of the harmonic oscillator is extensively studied in the literature. In the case of the NC space see  \cite{Halder:2016qwz}-\cite{Jahan:2012wh} for the recent works. The useful ingredient for this study in the partition function $Z^{k}(T, \varepsilon)$ depending with the degeneracy index $k$ as:
\begin{eqnarray}
Z^{k}(T, \varepsilon) = \sum_{n=0}^{\infty} (n+1) e^{- \beta E_{\varepsilon,n}^{k}}, \quad \beta = \frac{1}{k_B T},
\end{eqnarray}
$k_B$ is the 
Boltzmann constant.
The full partition function is the sum under all degeneracies terms as
$
Z(T, \varepsilon) = \sum_{k=0}^{n}  Z^{k}(T, \varepsilon).
$
From statistical mechanics point of view, the probability $p(i,j)$  of finding a system in a state $\ket{i,j}$ is given by the Boltzmann formula \cite{John}:
\bea
p(i,j)=\frac{e^{- \beta E_{\varepsilon,n}^{k}(i,j)}}{Z(T,\varepsilon)}.
\eea
Thereby the  others thermodynamic quantities such as the free enery $F=-\frac{1}{\beta}\log Z(T, \varepsilon) $, the entropy $S = - \frac{\partial F}{\partial T}$ , the internal energy $U = - \frac{\partial \ln Z}{\partial \beta}$, the heat capacity $C_v = \frac{\partial U}{\partial T}$ are  given  in  the table built in the figure \eqref{table}. 

\begin{figure}\begin{center}
{
\begin{tabular}{*{9}{|c|c} }
\hline  Quantities& Formulas\\
 \hline			
  Partition function & $Z(T, \varepsilon)  
= \frac{1}{4\sinh\big(\frac{\beta \hbar \omega}{2}\big)}\Big(1+\frac{m\omega^2\varepsilon^2\beta}{4}\coth^2\big(\frac{\beta \hbar \omega}{2}\big)\Big) +\mathcal{O}(\varepsilon^2)$ \\\hline
   Free energy& $F(T, \varepsilon) =\frac{2}{\beta}\log\Big[2 \sinh\big(\frac{\beta \hbar \omega}{2}\big)\Big]-\frac{m\omega^2\varepsilon^2}{4}\coth^2\big(\frac{\beta \hbar \omega}{2}\big) +\mathcal{O}(\varepsilon^2)$\\\hline
   Entropy& $S(T, \varepsilon)=-2 k_B\log \Big(2\sinh(\frac{\beta \hbar \omega}{2})\Big) + k_B \hbar \omega \beta \coth(\frac{\beta \hbar \omega}{2}) + \frac{1}{4} m k_B \hbar \omega^3 \beta^2 \varepsilon^2 \dfrac{\cosh(\frac{\beta \hbar \omega}{2})}{\sinh^3(\frac{\beta \hbar \omega}{2})}  + \mathcal{O}(\varepsilon^2)$ \\\hline
Internal energy&$U(T, \varepsilon) = \hbar \omega \coth(\frac{\beta \hbar \omega}{2})\Big[1 - \frac{m\omega^2 \varepsilon^2}{4} \Big(\frac{1}{\hbar \omega} \coth(\frac{\beta \hbar \omega}{2}) + \dfrac{\beta}{\sinh^2(\frac{\beta \hbar \omega}{2})}\Big)\Big] + \mathcal{O}(\varepsilon^2)$\\\hline 
Heat capacity&$C_v(T, \varepsilon) = - k_B \Big(\dfrac{\hbar \omega \beta/2}{\sinh^2(\frac{\hbar \omega \beta}{2})}\Big)^2 \Big[1 + m \omega^2 \varepsilon^2 \beta - (1 - \frac{1}{2} m \omega^2 \varepsilon^2 \beta) \cosh(\hbar \omega \beta)\Big]+ \mathcal{O}(\varepsilon^2)$\\
\hline 
 \end{tabular}
} \end{center}
\caption{The thermodynamic quantities of the lattice oscillator.}\label{table}
\end{figure}
\end{widetext}

We give the asymptotic behavior of the various functions at low temperatures $T<<\hbar\omega$.  First let us remark that if   $T$ is very small and goes to absolute $0$ we get
\bea
\frac{1}{\sinh\big(\frac{\beta \hbar \omega}{2}\big)}\rightarrow 0,\quad \frac{\beta \coth^2\big(\frac{\beta \hbar \omega}{2}\big)}{\sinh\big(\frac{\beta \hbar \omega}{2}\big)}\approx\frac{\beta}{e^{\frac{\beta \hbar \omega}{2}}}\rightarrow 0.
\eea
Then the partition function $Z(T,\varepsilon)$ goes to zero. This result is identical to what we obtain in the continuous space. On the other hand, for very high temperature and for a small value of the parameter $\varepsilon$,  $Z(T,\varepsilon)> Z(T,0)$ The variation $\Delta Z=Z(T,\epsilon)-Z(T,0)$ caracterize the correction due to the discretization of the space. As an example, let fix the constants $m$, $\omega$ and $\hbar$ to the unit. then for $T=10^5 K$ and $\varepsilon=10^{-11} $meter we get $\Delta Z=0.05.$

\begin{figure}
\begin{center}
\includegraphics[scale=0.4]{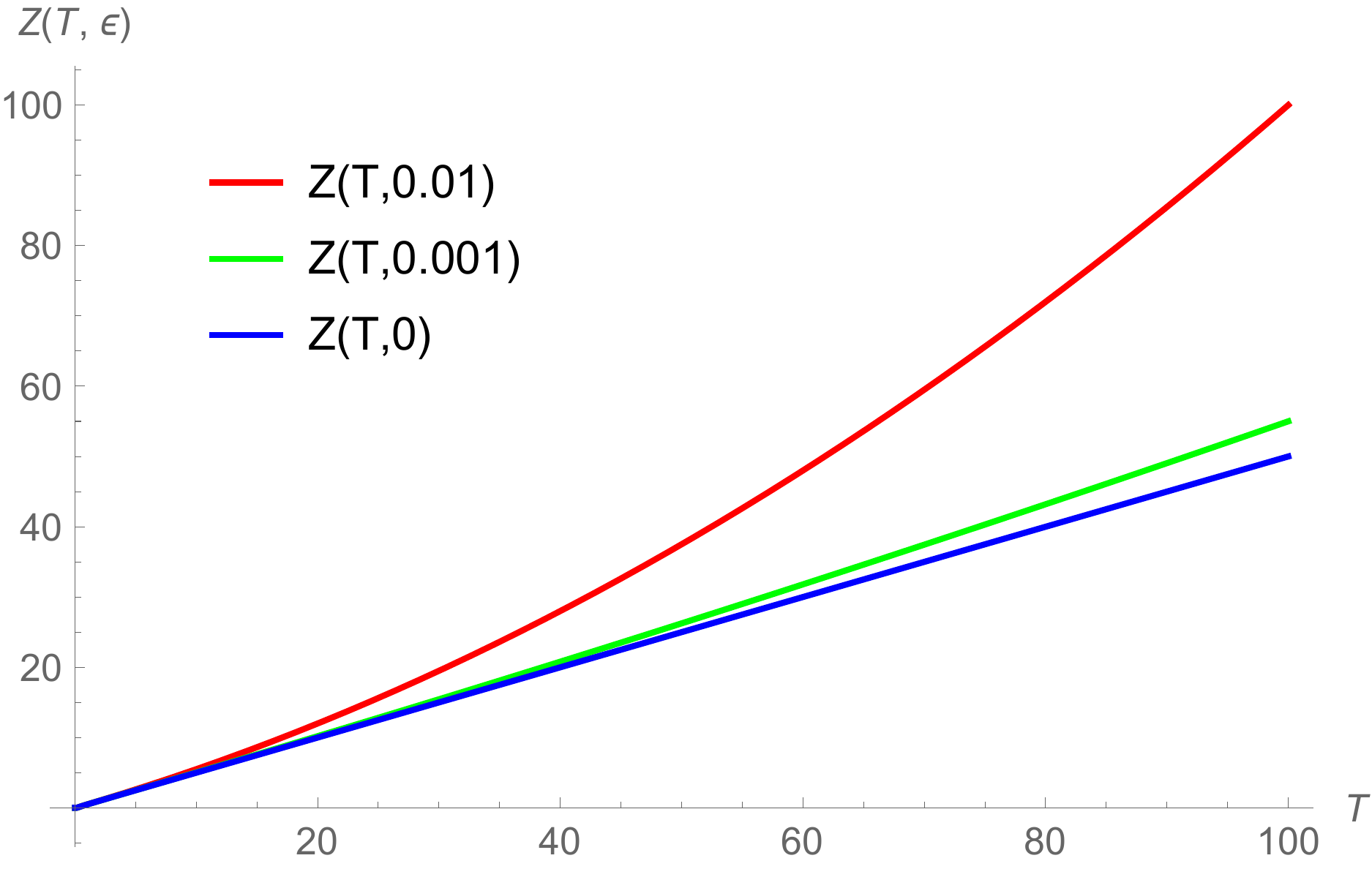}\quad\quad
\caption{Plot of the partition function for different values of  the lattice spacing $\varepsilon$.} \label{fig3} 
\end{center}
\end{figure}

\begin{figure}
\begin{center}
\includegraphics[scale=0.4]{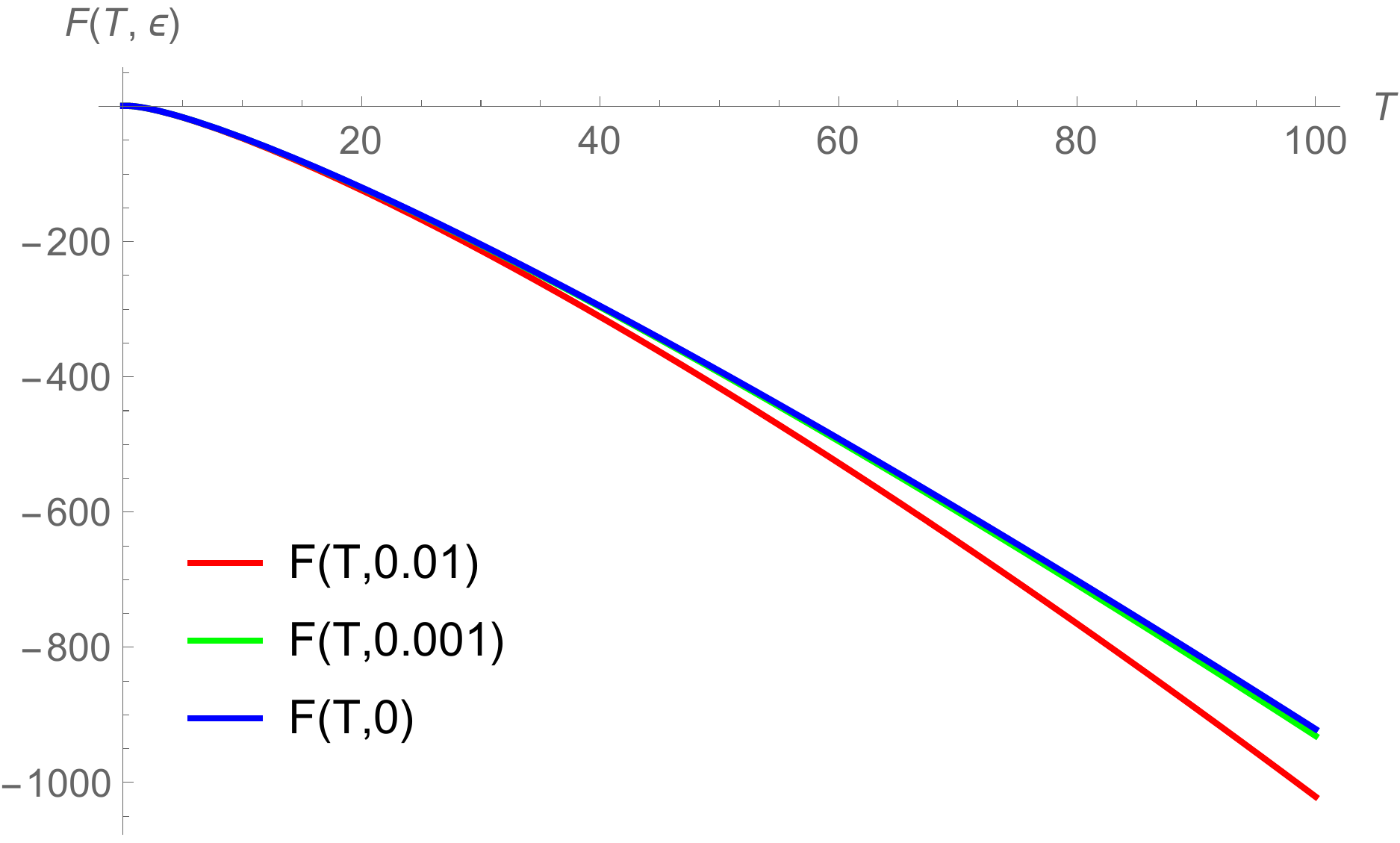}\quad\quad
\includegraphics[scale=0.4]{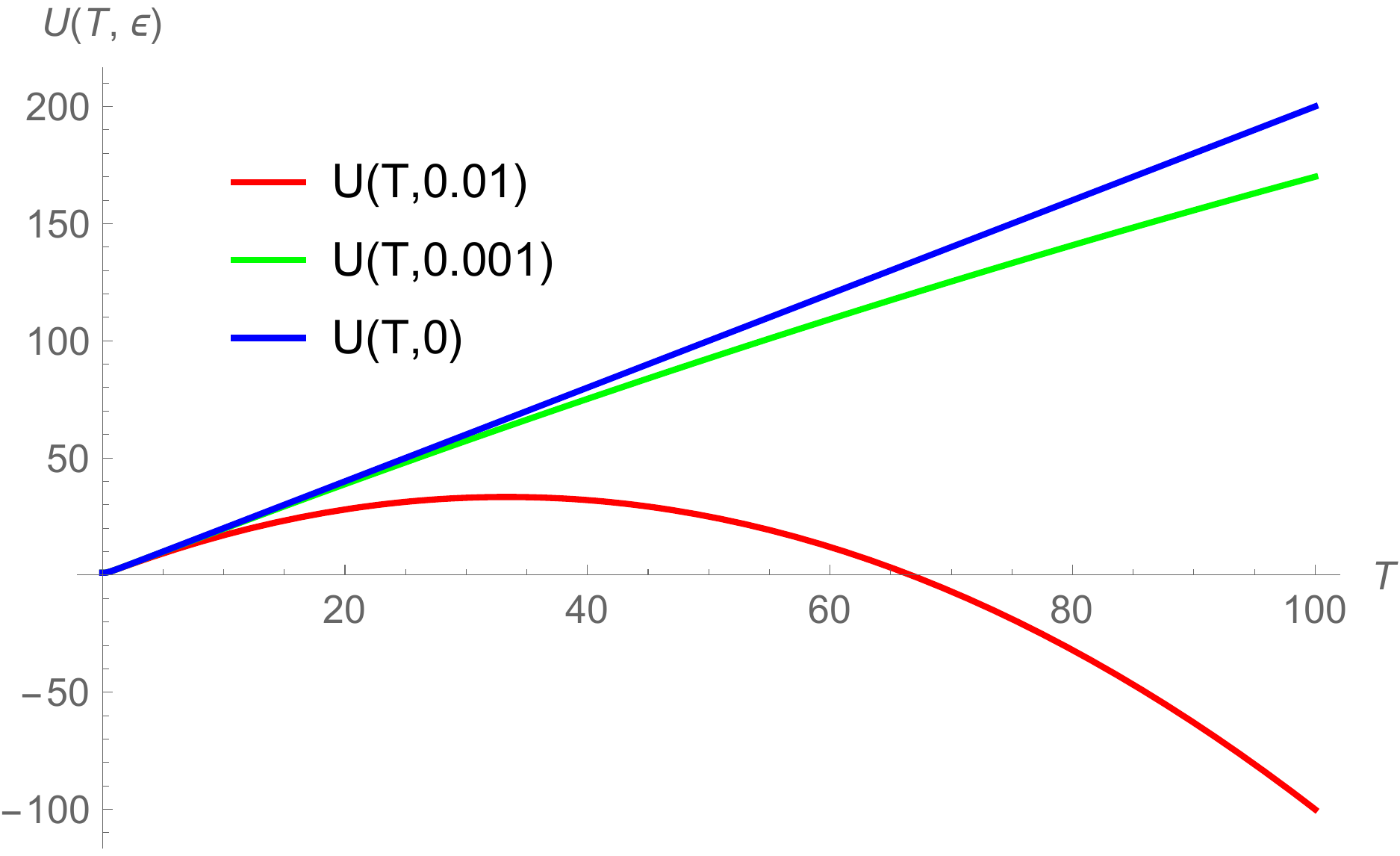}
\caption{Plot of the free energy and the internal energy for different values of  the lattice spacing $\varepsilon$. We find that for low temperatures, $\forall \varepsilon$  small the thermodynamic quantities coincide with the expected values when $\varepsilon =0$ (the continuum limit). The divergence appears for the high temperatures as we can easily see on these figures.} \label{fig4} 
\end{center}
\end{figure}

The free energy and the internal energy is reduced to
\bea\label{thermonew}
F(T<<\hbar\omega,\varepsilon)=U(T<<\hbar\omega,\varepsilon)\approx\hbar\omega-\frac{m\omega^2\varepsilon^2}{4}.
\eea
Remark that in the thermodynamic equilibrium applied to the harmonic oscillator, the free energy and internal energy are given by
\bea
F(T, 0) =\frac{2}{\beta}\log\Big[2 \sinh\Big(\frac{\beta \hbar \omega}{2}\Big)\Big],
\eea
\bea
U(T, 0) = \hbar \omega \coth\Big(\frac{\beta \hbar \omega}{2}\Big).
\eea
The limit $T<<\hbar\omega$ of these two quantities are reduced to $F(0,0)=\hbar\omega$ and $U(0,0)=\hbar\omega$,
which corresponds to the ground state energy of 2D quantum harmonic oscillator  and is in adequacy with thermodynamic equilibrium energy \cite{Halder:2016qwz}-\cite{Jahan:2012wh}. Therefore  the result \eqref{thermonew} is very close to what is obtained in thermodynamic quantum mechanics of the harmonic oscillator for the small value of $\varepsilon$ due to the fact that $\frac{m\omega^2\varepsilon^2}{4}$ remains small.  The contribution $\frac{m\omega^2\varepsilon^2}{4}$ appears as the first order correction of the lattice. On the other hand a divergence appears at high temperatures $T>>\hbar\omega$.   This behavior is illustrated in the figures \eqref{fig3} and \eqref{fig4}. The same analysis can be performed for the entropy $S$ and the heat capacity $C_v$.

\subsection{Harmonic oscillator in a noncommutative lattice}
In this subsection we consider the case where both momenta and coordinates are noncommutative, as presented   in the equations \eqref{ssss}. We will show that in particular case where  $\bar{\theta} = -m^2 \omega^2 \theta$, the eigenvalue problem can be determined as in the previous section. One of the most ambiguous aspect, namely in the case where $\bar{\theta} \neq -m^2 \omega^2 \theta$ is also considered and studied.
 The  Hamiltonian of the  oscillator
\bea
\hat{\mathfrak H}_\varepsilon =\frac{1}{2}[(\hat{P}_x^\varepsilon)^2+(\hat{P}_x^\varepsilon)^2]+\frac{1}{2}[(\hat{X}^\varepsilon)^2, (\hat{Y}^\varepsilon)^2],
\eea
 in the first order of $\varepsilon^2$ is split into:
\begin{eqnarray}
\hat{\mathfrak H}_\varepsilon = \hat{\mathfrak H}_0 + \varepsilon^2 \hat{\mathfrak W}+\mathcal{O}(\varepsilon^2)
\end{eqnarray}
where   $\hat{\mathfrak H}_0$ is assumed to be the nonperturbative  Hamiltonian and  $\hat{\mathfrak W}$ stands for the perturbation:
\bea\label{Hamildine}
\hat{\mathfrak H}_0 &= & \frac{\Omega^2}{2 m} (\hat{p}_x^2 + \hat{p}_y^2) + \frac{m \omega^2 \bar{\Omega}^2}{2} \Big((\hat{x}^\varepsilon)^2 + (\hat{y}^\varepsilon)^2 \Big) \cr
&&+ \frac{\widetilde{\Omega}}{m}(\hat{y}^\varepsilon \hat{p}_x -  \hat{x}^\varepsilon \hat{p}_y), 
\\
\hat{\mathfrak W} &=&  - \frac{1}{6 m \hbar^2} \Big[\Omega^2 (\hat{p}_x^4 + \hat{p}_y^4) + \tilde{\Omega} (\hat{y}^\varepsilon \hat{p}_x^3 - \hat{x}^\varepsilon  \hat{p}_y^3) \Big].
\eea
The  parameters  $\Omega$, $\bar\Omega$, $\widetilde\Omega$ are given by
\begin{eqnarray}
\Omega^2 &=& 1 + \frac{m^2 \omega^2 \theta^2}{4 \hbar^2}, \\
 \bar{\Omega}^2 &=& 1 + \frac{\bar{\theta}^2}{4 m^2 \hbar^2 \omega^2}, \\ \widetilde{\Omega} &=& \frac{\bar{\theta}}{2 \hbar} + \frac{m^2 \omega^2 \theta}{2 \hbar}.
\end{eqnarray}
The more general result concerning the eigenvalues problem of the Hamiltonian \eqref{Hamildine} can be obtained essentially by using the method described in the previous section. We construct the annihilation and creation operators
 ($\hat{\mathfrak b} $ and $\hat{\mathfrak b}^\dag $) as follows:
\begin{eqnarray}
&& \hat{\mathfrak b}_x = \frac{1}{\sqrt{2\Omega \bar{\Omega}\hbar m \omega}}\Big(i\Omega \hat{p}_x + m\omega \bar{\Omega} \hat{x}                   ^\varepsilon\Big), \\
&& \hat{\mathfrak b}_x^\dag = \frac{1}{\sqrt{2\Omega \bar{\Omega}\hbar m \omega}}\Big(- i\Omega \hat{p}_x + m\omega \bar{\Omega} \hat{x}^\varepsilon\Big) \\
&&  \hat{\mathfrak b}_y = \frac{1}{\sqrt{2\Omega \bar{\Omega}\hbar m \omega}}\Big(i\Omega \hat{p}_y + m\omega \bar{\Omega} \hat{y}^\varepsilon\Big), \\ &&\hat{\mathfrak b}_y^\dag = \frac{1}{\sqrt{2\Omega \bar{\Omega}\hbar m \omega}}\Big(- i\Omega \hat{p}_y + m\omega \bar{\Omega} \hat{y}^\varepsilon\Big).
\end{eqnarray}
They satisfy  the canonical commutation relation
$
\Big[\hat{\mathfrak b}_x,\hat{\mathfrak b}_x^\dag\Big] = \mathbb{1} = \Big[\hat{\mathfrak b}_y, \hat{\mathfrak b}_y^\dag\Big],
$
and according to which the Hamiltonian $\hat{\mathfrak H}_0$ can be factorized as follows:
\beq\label{Hamileee}
\hat{\mathfrak H}_0 =  \Omega \bar{\Omega}\hbar \omega\big(\hat{\mathfrak b}_x^\dag \hat{\mathfrak b}_x +  \hat{\mathfrak b}_y^\dag \hat{\mathfrak b}_y + \mathbb{1} \big) - i \frac{\widetilde{\Omega}\hbar}{m} \big(\hat{\mathfrak b}_x \hat{\mathfrak b}_y^\dag - \hat{\mathfrak b}_x^\dag\hat{\mathfrak b}_y \big).
\eeq
which corresponds to the Hamiltonian of two dimensional Landau problem in the symmetric gauge on NC space. Equivalently  the presence  of  magnetic fields  in  this  relation also suggest a NC structure for the
spacetime.  
The perturbation term is
\bea
\hat{\mathfrak W} &=& -\frac{\bar{\Omega} \omega}{24 \Omega} \Big[\Omega \bar{\Omega} m \omega \Big( (\hat{\mathfrak b}_x - \hat{\mathfrak b}_x^\dag)^4 + (\hat{\mathfrak b}_y - \hat{\mathfrak b}_y^\dag)^4\Big) \cr
&+& i \widetilde{\Omega} (\hat{\mathfrak b}_y + \hat{\mathfrak b}_y^\dag) (\hat{\mathfrak b}_x - \hat{\mathfrak b}_x^\dag)^3 - i \widetilde{\Omega}(\hat{\mathfrak b}_x + \hat{\mathfrak b}_x^\dag) (\hat{\mathfrak b}_y -\hat{\mathfrak b}_y^\dag)^3\Big].\cr
&&
\eea
For some purposes, it is useful to point out that the states of the form $\ket{n_x, n_y;0}$  defined in \eqref{eqeq2} may
 diagonalyze  the Hamiltonian \eqref{Hamileee}. Aiming at including the perturbation term $\hat{\mathcal W}$ in our diagonalization procedure let us consider the new operators:
\begin{eqnarray}
&& \hat{\mathfrak b}_+ = \frac{1}{\sqrt{2}}\big(\hat{\mathfrak b}_x + i \hat{\mathfrak b}_y\big), \quad \hat{\mathfrak b}_+^\dag = \frac{1}{\sqrt{2}}\big(\hat{\mathfrak b}_x^\dag - i \hat{\mathfrak b}_y^\dag\big), \\
&& \hat{\mathfrak b}_- = \frac{1}{\sqrt{2}}\big(
\hat{\mathfrak b}_x - i \hat{\mathfrak b}_y\big), \quad \hat{\mathfrak b}_-^\dag = \frac{1}{\sqrt{2}}\big(\hat{\mathfrak b}_x^\dag + i \hat{\mathfrak b}_y^\dag\big).
\end{eqnarray}
 Here also, the  canonical commutation relation is well satisfied i.e.: $
\Big[\hat{\mathfrak b}_+, \hat{\mathfrak b}_+^\dag\Big] = \mathbb{1} = \Big[\hat{\mathfrak b}_-,\hat{\mathfrak b}_-^\dag\Big] $. 
Then the Hamiltonian $\hat{\mathfrak H}_0$ and $\hat{\mathfrak W}$ become
\bea
\hat{\mathfrak H}_0 &=&  \Omega \bar{\Omega}\hbar \omega\Big(\hat{\mathfrak N}_+  + \hat{\mathfrak N}_- + \mathbb{1} \Big) - \frac{\widetilde{\Omega}\hbar}{m} \big(\hat{\mathfrak N}_- - \hat{\mathfrak N}_+\big) 
\\
\hat{\mathfrak W} &=&  - \frac{\bar{\Omega} \omega}{96\Omega} \Big[\Omega \bar{\Omega} m \omega \Big( A^{4} +B^{4}\Big) + \widetilde{\Omega} \Big(AC^{3} +BD^{3}\Big)\Big],\cr
&&
\eea
where
\bea
A=\hat{\mathfrak b}_+ -\hat{\mathfrak b}_+^\dag + \hat{\mathfrak b}_- - \hat{\mathfrak b}_-^\dag,\quad
B=\hat{\mathfrak b}_+ +\hat{\mathfrak b}_+^\dag - \hat{\mathfrak b}_- - \hat{\mathfrak b}_-^\dag\\
C=\hat{\mathfrak b}_+ - \hat{\mathfrak b}_+^\dag - \hat{\mathfrak b}_- +\hat{\mathfrak b}_-^\dag,\quad
D=\hat{\mathfrak b}_+ + \hat{\mathfrak b}_+^\dag + \hat{\mathfrak b}_- + \hat{\mathfrak b}_-^\dag
\eea
and  $\hat{\mathfrak N}_+ = \hat{\mathfrak b}_+^\dag \hat{\mathfrak b}_+$ and $\hat{\mathfrak N}_- =\hat{\mathfrak b}_-^\dag \hat{\mathfrak b}_-$ are the number operators.  Let $\{\ket{n_+, n_-;0}=\ket{n_+}\otimes\ket{n_-},n_-,n_+\in\mathbb{N}\}$ be a set of  Fock vectors such that
\begin{eqnarray}
\hat{\mathfrak b}_+^\dag \ket{n_+, n_-;0} = \sqrt{n_+ +1} \ket{n_+ + 1, n_-;0},\\
 \hat{\mathfrak b}_-^\dag \ket{n_+, n_-;0} = \sqrt{n_- +1} \ket{n_+, n_- + 1;0}
\end{eqnarray}
Then we get  
$$
\hat{\mathfrak N}_+ \ket{n_+, n_-;0} = n_+ \ket{n_+, n_-;0},$$ and
$$ \hat{\mathfrak N}_- \ket{n_+, n_-;0} = n_- \ket{n_+, n_-;0}.
$$
The states $\ket{n_+, n_-;0}$ solve the eigenvalue problem
\begin{equation}
\hat{\mathfrak H}_0 \ket{n_+, n_-;0} = {\mathcal E}_{0,n_+,n_-} \ket{n_+, n_-;0}
\end{equation}
with corresponding energies
\bea \label{Ett}
{\mathcal E}_{0,n_+,n_-}& =& \Omega \bar{\Omega}\hbar \omega\big(n_+ + n_- + 1 \big)  - \frac{\widetilde{\Omega}\hbar}{m} \big(n_- - n_+\big)\cr
&=& \Omega \bar{\Omega}\hbar \omega\big(n + 1 \big)  - \frac{\widetilde{\Omega}\hbar}{m} \jmath. 
\eea
where  $n= n_+ + n_-$ and $\jmath = n_- - n_+$. Concerning the perturbation $\hat{\mathfrak W}$, it seems that the states $ \ket{n_+, n_-;0} $ form a diagonalyzed basis, in  the case where
$\bar{\theta} = -m^2 \omega^2 \theta$, which refers to the solvable condition of the harmonic oscillator in noncommutative space when both momentum and coordinates are suppose to satisfy the nonvanish commutation relations.
The Hamiltonian $\hat{\mathfrak H}_0 $ and $\hat{\mathfrak W}$ become
\begin{eqnarray}
\hat{\mathfrak H}_0=   \Omega^2 \hat{H}_0,\quad
\hat{\mathfrak W}=  \Omega^2 \hat W 
\end{eqnarray}
where $\hat H$ 
 and $\hat W$ are defined in \eqref{samdino31}. Then, we find that the eigen-energy of $\hat{\mathfrak H}$ is
\begin{eqnarray}\label{spectrums}
&&{\mathcal E}_{\varepsilon,n}^{k}= \Omega^2\Big[ \bar{\mathcal E}_{0n}- \frac{ m \omega^2 \varepsilon^2}{4} \Big(n^2 + n(1 - 2 k) + 2 k^2 + 1 \Big)\Big].\cr
&&
\end{eqnarray}
with $\bar{\mathcal E}_{0n}=\hbar \omega (n + 1)$ and $k = 0, 1, ..., n$ are the order of degeneracy. Let us note that all the thermodynamic properties derived in the last subsection can also be performed using the spectrum \eqref{spectrums} and we arrive at the same conclusion. 

\begin{widetext}
Now we will focus on the case of arbitrary positive values of the parameters $\theta$ and $\bar\theta$. 
The first order correction energy ${\mathcal E}_{n_+,n_-}^{1k}$ is obtained by the following system
\beq\label{gregorie}
\sum_{\substack{j=0 \\ j \neq k}}^{n} c_j \; {}^k\langle \hat{\mathfrak W} \rangle^j+  c_k\big({}^k\langle \hat{\mathfrak W}\rangle^k -  {\mathcal E}_{n_+,n_-}^{1k}\big) = 0,
\eeq
where the matrix elements ${}^k\langle \hat{\mathfrak W} \rangle^j:= {}^k\bra{0;n_+,n_-}\hat{\mathfrak W}\ket{n_+, n_-;0}^j$ are explicitly given by
\begin{eqnarray}
{}^k\langle \hat{\mathfrak W} \rangle^j = \left \{ \begin{array}{l}
 - \frac{\bar{\Omega}^2 m \omega^2}{8} \Big((n+1)(n+2) +2 k n -2 k^2\Big) 
-  \frac{\widetilde{\Omega} \bar{\Omega} \omega}{8\Omega} (n+1)(n-2k), \quad  j = k, 
\cr
- \frac{\bar{\Omega}^2 m \omega^2}{8} \sqrt{k(k-1)(n-k+1)(n-k+2)}, \quad j = k-2, 
\cr
- \frac{\bar{\Omega}^2 m \omega^2}{8} \sqrt{(k+1)(k+2)(n-k-1)(n-k)}, \quad  j= k+2, 
\cr
0  \quad \textrm{otherwise}
\end{array} \right .
\end{eqnarray}
Our interest is the determinant of the matrix
 $({\mathfrak G})$ similar to expression  \eqref{Gmat}:
\begin{eqnarray}\label{matnew25}
({\mathfrak G}) : \left \{ \begin{array}{l}
{\mathfrak G}_{k,k} = - \frac{\bar{\Omega}^2 m \omega^2}{8} \Big((n+1)(n+2) +2 k n -2 k^2\Big) -  \frac{\widetilde{\Omega} \bar{\Omega} \omega}{8\Omega} (n+1)(n-2k) -  {\mathcal E}_{n}^{1k}, \cr
{\mathfrak G}_{k,k-2} =- \frac{\bar{\Omega}^2 m \omega^2}{8} \sqrt{k(k-1)(n-k+1)(n-k+2)}, \cr
{\mathfrak G}_{k,k+2} = - \frac{\bar{\Omega}^2 m \omega^2}{8} \sqrt{(k+1)(k+2)(n-k-1)(n-k)}, \cr
0  \quad \textrm{otherwise}
\end{array} \right .
\end{eqnarray}
The matrix $\mathfrak G$ is not diagonal but symmetric, i.e.
 ${\mathfrak G}_{k,k+2}={\mathfrak G}_{k+2,k}  $. The solution of  equation $\det\mathfrak G=0$ can no longer be obtained by direct calculation for arbitrary value of the integer $n$. We provide here this solution order by order to this quantum number $n$. So the first order  corrections of the energy spectrum become
\bea
n = 0 : \quad \,\quad  {\mathcal E}_{0}^{10}=-\frac{\bar{\Omega}^2 m \omega^2}{4}
\eea
\bea
n = 1 : \left \{ \begin{array}{l}
 {\mathcal E}_{1}^{11}=- \frac{3\bar{\Omega}^2 m \omega^2}{4} + \frac{\widetilde{\Omega} \bar{\Omega} \omega}{4\Omega}\\
 {\mathcal E}_{1}^{10}=- \frac{3\bar{\Omega}^2 m \omega^2}{4} -  \frac{\widetilde{\Omega} \bar{\Omega} \omega}{4\Omega}\end{array} \right .
\eea
\bea
n = 2 : \left \{ \begin{array}{l}
{\mathcal E}_{2}^{12}= - \frac{3 \bar{\Omega}^2 m \omega^2}{2} + \frac{\omega \bar{\Omega}}{4 \Omega}\sqrt{9 \widetilde{\Omega}^2 + \Omega^2 \bar{\Omega}^2 m^2 \omega^2}\\
{\mathcal E}_{2}^{11}=- \frac{7\bar{\Omega}^2 m \omega^2}{4}\\
{\mathcal E}_{2}^{10}=- \frac{3 \bar{\Omega}^2 m \omega^2}{2} - \frac{\omega \bar{\Omega}}{4 \Omega}\sqrt{9 \widetilde{\Omega}^2 + \Omega^2 \bar{\Omega}^2 m^2 \omega^2}
\end{array} \right . 
\eea
\begin{eqnarray}
n = 3 : \left \{ \begin{array}{l}
{\mathcal E}_{3}^{13}= \frac{2 \bar{\Omega} \widetilde{\Omega} \omega - 11 \Omega \bar{\Omega}^2 m \omega^2 }{4 \Omega} + \frac{\bar{\Omega} \omega}{2 \Omega } \sqrt{4 \widetilde{\Omega}^2 + 2 \Omega \bar{\Omega} \widetilde{\Omega} m \omega +\Omega^2 \bar{\Omega}^2 m^2 \omega^2}\\
{\mathcal E}_{3}^{12}=-  \frac{2 \bar{\Omega} \widetilde{\Omega} \omega + 11 \Omega \bar{\Omega}^2 m \omega^2 }{4 \Omega} + \frac{\bar{\Omega} \omega}{2 \Omega } \sqrt{4 \widetilde{\Omega}^2 - 2 \Omega \bar{\Omega} \widetilde{\Omega} m \omega +\Omega^2 \bar{\Omega}^2 m^2 \omega^2} \\
{\mathcal E}_{3}^{11} = \frac{2 \bar{\Omega} \widetilde{\Omega} \omega - 11 \Omega \bar{\Omega}^2 m \omega^2 }{4 \Omega} - \frac{\bar{\Omega} \omega}{2 \Omega } \sqrt{4 \widetilde{\Omega}^2 + 2 \Omega \bar{\Omega} \widetilde{\Omega} m \omega +\Omega^2 \bar{\Omega}^2 m^2 \omega^2}\\
{\mathcal E}_{3}^{10}=-  \frac{2 \bar{\Omega} \widetilde{\Omega} \omega + 11 \Omega \bar{\Omega}^2 m \omega^2 }{4 \Omega} - \frac{\bar{\Omega} \omega}{2 \Omega } \sqrt{4 \widetilde{\Omega}^2 - 2 \Omega \bar{\Omega} \widetilde{\Omega} m \omega +\Omega^2 \bar{\Omega}^2 m^2 \omega^2}
\end{array} \right.
\end{eqnarray}
For $n\geq 4$ the computation of the determinant of $\mathfrak G$  leads to the mixing of real and complex values as solutions of equation  \eqref{gregorie}. The complex energies  cannot be taking into account in our analysis. Moreover for the moment we have no method to classify these solutions. Then we consider only the quantum numbers $n=0,1,2,3$ as given above. 
The correction of the states $\ket{N}$ namely $\ket{n_+, n_-;1}$ are given 
using the first order perturbation equation:
\begin{eqnarray}\label{saline}
&& \ket{n_+, n_-;1} = - \sum_{\substack{\ell_+, \ell_- \\ \ell_+ \notin \mathcal{D} \\ \ell_- \notin \mathcal{D'}}} \sum_{j=0}^{n} c_j \frac{ \bra{0;\ell_+,\ell_-}\hat{\mathfrak W}\ket{n_+, n_-;0}^j}{{\mathcal E}_{\ell_+,\ell_-}^0 - {\mathcal E}_{n_+,n_-}^0} \ket{\ell_+,\ell_-;0},
\end{eqnarray}
where $\mathcal{D} = \{n_++n_-- j ;\; 0 \le j \le n_++n_-\},\,\,\mathcal{D'} = \{j;\;  0 \le j \le n_++n_-\}$. 
Let $\bra{0;\ell_+,\ell_-}\hat{\mathfrak W}\ket{n_+, n_-;0}^j :=\langle \hat{\mathfrak W}\rangle_{\ell n}^j$
we can determine  $\ket{n_+, n_-;1}$ by replacing in \eqref{saline} the following relation
\begin{eqnarray}
 \langle \hat{\mathfrak W}\rangle_{\ell n}^j=  \left\{ 
\begin{array}{ll}
- \frac{\bar{\Omega}^2 m \omega^2}{8} \sqrt{j(j-1)(n-j)(n-j-1)} & \ell_- = j-2; \ell_+ =n-j-2 \\
- \frac{\bar{\Omega}^2 m \omega^2}{8} \sqrt{(j+1)(j+2)(n-j+1)(n-j+2)} & \ell_- = j+2 ; \ell_+ =n-j+2  \\
 (\frac{\widetilde{\Omega} \bar{\Omega} \omega}{8\Omega} (n - 2 j) + \frac{\bar{\Omega}^2 m \omega^2}{4} (n+2)) \sqrt{(j+1)(n-j+1)} &  \ell_- = j+1 ;\ell_+ =n-j+1 \\
(\frac{\bar{\Omega}^2 m \omega^2}{12} - \frac{\widetilde{\Omega} \bar{\Omega} \omega}{24\Omega}) \sqrt{(j+1)(j+2)(j+3)(n-j)} &  \ell_- = j+3 ; \ell_+ =n-j-1 \\
(-\frac{\widetilde{\Omega} \bar{\Omega} \omega}{4\Omega} j + (\frac{\bar{\Omega}^2 m \omega^2}{4} + \frac{\widetilde{\Omega} \bar{\Omega} \omega}{8\Omega}) n) \sqrt{j(n-j)} & \ell_- = j-1; \ell_+ =n-j-1 \\
 (\frac{\bar{\Omega}^2 m \omega^2}{12} - \frac{\widetilde{\Omega} \bar{\Omega} \omega}{24\Omega}) \sqrt{j(j-2)(j-1)(n-j+1)} & \ell_- = j-3; \ell_+ =n-j+1 \\
 (\frac{\bar{\Omega}^2 m \omega^2}{12} + \frac{\widetilde{\Omega} \bar{\Omega} \omega}{24\Omega})  \sqrt{(j+1)(n-j)(n-j-1)(n-j-2)} & \ell_- = j+1 ; \ell_+=n-j-3 \\
(\frac{\bar{\Omega}^2 m \omega^2}{12} + \frac{\widetilde{\Omega} \bar{\Omega} \omega}{24\Omega}) \sqrt{j(n-j+1)(n-j+2)(n-j+3)} & \ell_- = j-1; \ell_+ =n-j+3
\end{array} \right .
\end{eqnarray}
\,\,For $n=0$ we get:
\begin{eqnarray}
\ket{0, 0;1} &=& \frac{1}{\sqrt{17}} \ket{2, 2;0} - \frac{4}{\sqrt{17}} \ket{1,1;0}.
\end{eqnarray}
\,\,For $n= 1$, we get
\begin{eqnarray}
\ket{n_+, n_-;1} &=&  c_0  \Big[\frac{\bar{\Omega}^2 m^2 \omega^2}{8 \hbar (2 \Omega  \bar{\Omega} m \omega +\widetilde{\Omega} n_-)} \sqrt{3} \, \ket{3,2;0} - \frac{\widetilde{\Omega} \bar{\Omega} m \omega + 6\Omega \bar{\Omega}^2 m^2 \omega^2}{16\Omega \hbar (\Omega  \bar{\Omega} m \omega + \widetilde{\Omega} n_-)}\sqrt{2} \, \ket{2,1;0} \cr
&-& \frac{2\Omega \bar{\Omega}^2 m^2 \omega^2 + \widetilde{\Omega} \bar{\Omega} m \omega}{48\Omega \hbar (\Omega  \bar{\Omega} m \omega + \widetilde{\Omega} (n_- - 2))} \sqrt{6} \, \ket{0,3;0}\Big] + c_1  \Big[\frac{\bar{\Omega}^2 m^2 \omega^2}{8 \hbar (2 \Omega  \bar{\Omega} m \omega +\widetilde{\Omega} (n_- - 1))} \sqrt{3} \, \ket{2, 3;0} \cr
&-& \frac{-\tilde{\Omega} \bar{\Omega} m \omega + 6\Omega \bar{\Omega}^2 m^2 \omega^2}{16\Omega \hbar (\Omega  \bar{\Omega} m \omega + \widetilde{\Omega} (n_- - 1))}\sqrt{2} \, \ket{1,2;0} - \frac{2\Omega \bar{\Omega}^2 m^2 \omega^2 + \widetilde{\Omega} \bar{\Omega} m \omega}{48 \Omega \hbar (\Omega \bar{\Omega} m \omega + \widetilde{\Omega} (n_- +1))} \sqrt{6} \, \ket{3,0;0} \Big]\cr
&&
\end{eqnarray}
where $n_+ = 1$ and $n_- = 0$  or $n_+ = 0$ and $n_- = 1$. The constants $c_0$ and $c_1$ are determined using the normalization conditions
$\braket{1;1, 0}{1, 0;1} = 1$ and 
$\braket{1;0, 1}{0, 1;1} = 1$.\\
\end{widetext}

Let us now discuss the consistency of our result. First of all we show  the difficulties around  the direct computation of the eigen-equation of the harmonic oscillator in 2D lattice not only in ordinary space but also in NC space. These difficulties come from the fact that the corresponding lattice analogue of creation and annihilation operators are not Ladder operators. We provide using perturbative method, the solution of this complicate differential equation. Despite this very promising result the case of NC space need to be revisited by defined the new basis which can help to diagonalize the matrix $({\mathfrak G})$  of expression \eqref{matnew25} similar to expression  \eqref{Gmat}. Finally  the general solution i.e. the case where $\bar{\theta} \neq -m^2 \omega^2 \theta$ need to be also scrutinized.

\section{ Conclusion and remarks}\label{sec4}
In this paper we have solved the harmonic oscillator in the  $2d$ lattice. First we have considered the case of ordinary quantum mechanics.  We showed that the direct computation of the eigenvalues by using the analogue of  Ladder operator is not satisfactory as far as the physical relevance is concerned, due to the appearance of coordinates dependency in the energies. Also the continuous limit i.e. $\varepsilon=0$ is not well satisfied. The first order approximation of the lattice spacing $\varepsilon$ has been considered and the perturbation computation of the energy spectrum given. The statistical thermodynamic properties of the model have also been given. On the other hand the same question is addressed for general noncommutativity between coordinates and momenta. 
We have come to the conclusion that the eigenvalue problem may be solved in the case where $\bar{\theta} = -m^2 \omega^2 \theta$. The more general case where this relation is not satisfied has also been examined. We hope that it will be possible to construct a new Fock states in which the matrix  $\hat{\mathfrak W}$ maybe diagonalizable. This question deserve to be addressed and will be considered  in forthcoming work.

 \section*{Acknowledgements} 
D.O.S research at the Max-Planck Institute is supported by the Alexander von Humboldt foundation. S.L.G thanks the Max-Planck Institute for invitation and financial support.


\end{document}